\documentclass[aps,prev,twocolumn,preprintnumbers,floatfix,nofootinbib]{revtex4-1}
\pdfoutput=1
\usepackage[english]{babel}
\usepackage{graphicx,color}
\usepackage{bm}
\usepackage{times}
\usepackage{slashed}
\usepackage{multirow}
\usepackage[usenames,dvipsnames,svgnames,table]{xcolor}
\usepackage{adjustbox}
\usepackage{slashed}
\usepackage{booktabs}
\usepackage{makecell}

\usepackage{amsthm}
\usepackage{amssymb,amsmath}
\usepackage{subfigure}
\usepackage{hyperref}
\usepackage[dvipsnames]{xcolor}
\definecolor{bluencs}{rgb}{0.0, 0.53, 0.74}
\definecolor{darkcyan}{rgb}{0.0, 0.55, 0.55}
\definecolor{hanblue}{rgb}{0.27, 0.42, 0.81}
\hypersetup{
	linktocpage=true,
	setpagesize=true,
	urlcolor=blue,
	citecolor=blue,
	linkcolor=blue, 
	menucolor=cyan,
	colorlinks=true, 
	filecolor=magenta,
	citebordercolor=blue,
	pdftitle={Single Production of a Vector-Like Top as a Probe of Charged Higgs Bosons at a Muon-Proton Collider},
	pdfsubject={Latex},
	pdfauthor={Bk},
	pdfkeywords={BSM},
	unicode=true
}
\usepackage[capitalize]{cleveref}

\newcommand{\be}{\begin{equation}}
\newcommand{\ee}{\end{equation}}
\newcommand{\bea}{\begin{eqnarray}}
\newcommand{\eea}{\end{eqnarray}}

\usepackage{xfrac}
 
\usepackage{float}
\usepackage{bbding,pifont}
\definecolor{brilliantrose}{rgb}{1.0, 0.33, 0.64}
\definecolor{lawngreen}{rgb}{0.49, 0.99, 0.0}
\definecolor{magenta}{rgb}{1.0, 0.0, 1.0}
\makeatletter
\let\old@float\@float
\def\@float{\let\centering\relax\old@float}
\makeatother
\begin{document}

\title{Single Production of a Vector-Like Top as a Probe of  Charged Higgs Bosons at a Muon-Proton Collider}

\author{R. Benbrik$^{1}$}
\email{r.benbrik@uca.ac.ma}
\author{M. Berrouj$^{1}$}
\email{mbark.berrouj@ced.uca.ma}
\author{M. Boukidi$^{2}$}
\email{mohammed.boukidi@ifj.edu.pl}
\author{H. Chatoui$^{3}$}
\email{chatouih@yahoo.fr}
\author{M. Ech-chaouy$^{1}$}
\email{m.echchaouy.ced@uca.ac.ma}
\author{K. Kahime$^{4}$}
\email{Kahimek@gmail.com}
\author{K. Salime$^{1}$}
\email{k.salime.ced@uca.ac.ma}

\affiliation{$^1$ Laboratory of Physics, Energy, Environment, and Applications, Cadi Ayyad University,
	Sidi Bouzid, P.O. Box 4162, Safi, Morocco.}
\affiliation{$^2$ Institute of Nuclear Physics, Polish Academy of Sciences, ul. Radzikowskiego 152, Cracow, 31-342, Poland.}
\affiliation{$^3$ ISPITS-Marrakech, \& Private University of Marrakech (UPM) Morocco.}

\affiliation{$^4$ Laboratoire Interdisciplinaire de Recherche en Environnement, Management, Energie et Tourisme (LIREMET), ESTE, Cadi Ayyad University, B.P. 383, Essaouira, Morocco.}

\begin{abstract}
	We investigate the discovery prospects for a vector-like top partner ($T$) within the Type-II Two-Higgs-Doublet Model (2HDM-II) at future high-energy $\mu p$ colliders. The analysis focuses on the charged Higgs decay mode $\mu^+ p \to \nu_\mu\, \bar{b}T \to \nu_\mu\, \bar{b}H^+b$, with the subsequent decay $H^+\to t\bar{b}$ yielding a final state with multiple $b$-jets and a charged lepton. A detailed detector-level simulation is carried out for benchmark configurations with charged Higgs masses around 600-650~GeV and vector-like top masses in the range $m_T \simeq 1.0$-$1.8$~TeV. For an integrated luminosity of 100~fb$^{-1}$, discovery significances above $5\sigma$ are obtained across several benchmark points, remaining robust against systematic uncertainties up to 20\%. At higher luminosities of 234~fb$^{-1}$, the sensitivity exceeds $10\sigma$ for the lightest benchmarks and stays above $5\sigma$ even in the presence of 30\% systematics.
\end{abstract}

\maketitle

\section{Introduction}

The discovery of the Higgs boson by the ATLAS and CMS collaborations at the Large Hadron Collider (LHC)~\cite{ATLAS:2012yve,CMS:2012qbp} confirmed the Higgs mechanism as the source of electroweak symmetry breaking and mass generation in the Standard Model (SM). While this achievement completed the SM particle spectrum, it also left key questions about the nature of the Higgs sector unanswered. These open issues have motivated the study of extensions of the scalar sector that can accommodate new dynamics or explain observed patterns in fermion masses and mixing. Among the simplest possibilities, the Two-Higgs-Doublet Model (2HDM)~\cite{Branco:2011iw,Draper:2020tyq} introduces a second scalar doublet, leading to a richer Higgs spectrum that includes two neutral states ($H$, $A$) and a charged Higgs boson ($H^\pm$). The discovery of these additional Higgs particles would mark clear evidence for physics beyond the SM.

A natural and well-motivated extension of the 2HDM is obtained by adding vector-like quarks (VLQs)~\cite{Aguilar-Saavedra:2013qpa,Buchkremer:2013bha,Fuks:2016ftf,Alves:2023ufm,Gopalakrishna:2013hua,Han:2023ied,Han:2025itd,Han:2022jcp,Yang:2022wfa,Tian:2021oey,Yang:2021dtc,Cao:2022mif,Yang:2024aav,Banerjee:2024zvg,Banerjee:2023upj,Vignaroli:2012si,Vignaroli:2015ama,Moretti:2025ckw,Han:2023ied,Han:2025itd,Yang:2022wfa,Yang:2021dtc,Cao:2022mif,Banerjee:2024zvg,Cui:2022hjg}. These heavy fermions have left- and right-handed components that transform identically under the SM gauge group, allowing gauge-invariant mass terms that are independent of electroweak symmetry breaking. Such particles arise in many theoretical frameworks, including Randall-Sundrum models~\cite{Randall:1999ee,Carena:2007tn,Gopalakrishna:2011ef}, $E_6$-based grand unified theories~\cite{Hewett:1988xc}, Little Higgs models~\cite{Arkani-Hamed:2002ikv,Schmaltz:2005ky}, and composite Higgs scenarios~\cite{Dobrescu:1997nm,Hill:2002ap,Agashe:2004rs,Barbieri:2007bh}. When embedded in the 2HDM, VLQs interact with the extended Higgs sector through new Yukawa couplings that can lead to non-standard decay patterns and striking collider signatures~\cite{Arhrib:2024tzm,Arhrib:2024dou,Arhrib:2024mbq,Benbrik:2022kpo,Abouabid:2023mbu,Benbrik:2023xlo,Benbrik:2024bxt,Aguilar-Saavedra:2017giu,Dermisek:2019vkc,Dermisek:2021zjd,Dermisek:2020gbr,Ghosh:2023xhs}.

In this framework, decay channels such as $T/B \to H^\pm b/t$, $T/B \to H t/b$, and $T/B \to A t/b$ may dominate over the conventional SM-like modes ($Wq$, $Zq$, $hq$). These new possibilities substantially change the expected branching ratios of VLQs and can relax constraints derived from standard searches~\cite{,Benbrik:2025kvz}. As demonstrated in our previous studies~\cite{Arhrib:2024tzm,Arhrib:2024dou,Arhrib:2024mbq,Benbrik:2022kpo,Benbrik:2023xlo,Benbrik:2024bxt,Abouabid:2023mbu,Benbrik:2025nfw,Benbrik:2024hsf,Arhrib:2024nbj}, such scenarios within the 2HDM-II can modify the electroweak and flavor limits, provide viable parameter regions consistent with current data, and open new channels for collider discovery.

The ATLAS and CMS collaborations have carried out extensive searches for VLQs through both single and pair production processes~\cite{ATLAS:2024fdw,CMS:2024bni}. So far, no evidence has been observed, and mass limits typically exceed 1.3-1.5~TeV depending on the VLQ representation and decay mode. The absence of signals in the standard final states may indicate that VLQs preferentially decay into non-SM particles, such as extra Higgs bosons, which are not yet systematically explored. This motivates dedicated analyses of models like the 2HDM+VLQ, where new decays into charged or neutral Higgs states can dominate.

In this Letter, we study the discovery potential of a vector-like top partner ($T$) in the Type-II 2HDM at a high-energy $\mu p$ collider. We focus on the charged Higgs channel 
$\mu^+p \to \nu_\mu\,\bar{b}T \to \nu_\mu\,\bar{b}H^+b$, 
with the subsequent decay $H^+ \to t\bar{b}$ producing a final state with multiple $b$-jets and a charged lepton. Owing to their high center-of-mass energy and clean environment, $\mu p$ colliders~\cite{Kaya:2018kyt,Kaya:2019ecf,Ketenoglu:2022fzo,Akturk:2024evo,Akturk:2024evo} provide an excellent opportunity to probe heavy fermions and extended Higgs sectors beyond the reach of the LHC.  

Our detector-level analysis demonstrates that, for charged Higgs masses around 600-650 GeV, a $5\sigma$ discovery can be achieved for $T$ masses up to about 1.8 TeV with $\mathcal{L}=100~\text{fb}^{-1}$, remaining robust against systematic uncertainties up to 20\%. At $\mathcal{L}=234~\text{fb}^{-1}$, significances exceed $10\sigma$ across all benchmark scenarios. 

The paper is organized as follows: In Section~\ref{sec-F}, we introduce the 2HDM-II+VLQs framework and outline the simulation setup for production and decay processes. Section~\ref{sec-A} presents the numerical results. Finally, Section~\ref{conclusion} summarizes our findings and discusses their implications for future collider experiments.

\section{Framework}
\label{sec-F}

We outline here the essential ingredients of the 2HDM-II extended by vector-like quarks (VLQs). The scalar sector corresponds to a $\mathcal{CP}$-conserving Two-Higgs-Doublet Model with two $SU(2)_L$ doublets, $\Phi_1$ and $\Phi_2$, obeying a softly broken $\mathbb{Z}_2$ symmetry ($\Phi_1 \to -\Phi_1$) introduced to control flavor violation~\cite{Branco:2011iw,Gunion:1989we}. The scalar potential reads
\begin{align}
\mathcal{V} &= m^2_{11}\Phi_1^\dagger\Phi_1+m^2_{22}\Phi_2^\dagger\Phi_2
-\big(m^2_{12}\Phi_1^\dagger\Phi_2+{\rm h.c.}\big)  \nonumber \\
&+\tfrac{1}{2}\lambda_1(\Phi_1^\dagger\Phi_1)^2+\tfrac{1}{2}\lambda_2(\Phi_2^\dagger\Phi_2)^2
+\lambda_3(\Phi_1^\dagger\Phi_1)(\Phi_2^\dagger\Phi_2) \nonumber\\
&+\lambda_4(\Phi_1^\dagger\Phi_2)(\Phi_2^\dagger\Phi_1)
+\Big[\tfrac{1}{2}\lambda_5(\Phi_1^\dagger\Phi_2)^2+{\rm h.c.}\Big].
\label{pot}
\end{align}
All parameters are real. After electroweak symmetry breaking, the doublets can be rotated into a Higgs basis where only one acquires a vacuum expectation value (VEV). The scalar sector can be conveniently described by the independent set
$\{ m_h, m_H, m_A, m_{H^\pm}, \tan\beta=v_2/v_1, \sin(\beta-\alpha), m_{12}^2\}$,
where $\tan\beta$ is the ratio of the two VEVs and $m_{12}^2$ controls the soft $\mathbb{Z}_2$ breaking.

Tree-level flavor-changing neutral currents are avoided by extending the $\mathbb{Z}_2$ symmetry to the Yukawa sector, giving four possible types of fermion couplings. In the Type-II realization considered here, $\Phi_2$ couples to up-type quarks while $\Phi_1$ couples to down-type quarks and charged leptons. We work in the alignment limit~\cite{Draper:2020tyq}, where the lightest scalar state reproduces the observed 125~GeV Higgs boson.

\vspace{3pt}
\noindent\textbf{Vector-like quark sector.}  
VLQs are added in representations that allow gauge-invariant mass terms. The possible multiplets under $SU(2)_L$ are
\begin{align}
& T_{L,R}^0  && \text{(singlet)}, \notag\\
& (X\,T^0)_{L,R},\; (T^0\,B^0)_{L,R} && \text{(doublets)}, \notag\\
& (X\,T^0\,B^0)_{L,R},\; (T^0\,B^0\,Y)_{L,R} && \text{(triplets)}.
\end{align}
The superscript “0” labels weak-interaction eigenstates. Their electric charges are
$Q_X=5/3$, $Q_T=2/3$, $Q_B=-1/3$, and $Q_Y=-4/3$,
with $T$ and $B$ sharing the same charges as the SM top and bottom quarks.

In the presence of new weak multiplets, the physical quark eigenstates become mixtures of the SM and VLQ components, inducing small modifications to $Z$-boson couplings. Precision data and atomic parity-violation measurements constrain these mixings strongly for light quarks but allow moderate values for the third generation~\cite{ParticleDataGroup:2012pjm}. Consequently, only mixing with the top and bottom quarks is retained in our setup.

\vspace{3pt}
\noindent\textbf{Yukawa structure and mixing.}
In the $\mathbb{Z}_2$ basis, the Yukawa Lagrangian for the three SM generations and one VLQ doublet $Q=(T^0,B^0)^T$ is
\begin{align}
\mathcal{L}_Y &=
-y_{ij}^u\,\bar q_{Li}\tilde{\Phi}_2 u_{Rj}
-y_{ij}^d\,\bar q_{Li}\Phi_1 d_{Rj}  \nonumber\\
&-y_{4j}^u\,\bar Q_L\tilde{\Phi}_2 u_{Rj}
-y_{4j}^d\,\bar Q_L\Phi_1 d_{Rj} +{\rm h.c.},
\end{align}
where $\tilde{\Phi}_2=i\sigma_2\Phi_2^*$.
We assume that the heavy doublet mixes predominantly with the third generation, consistent with the observed mass hierarchy and flavor constraints~\cite{Aguilar-Saavedra:2013wba,Barger:1995dd,Frampton:1999xi,Aguilar-Saavedra:2002phh}.  

After electroweak symmetry breaking, the mass terms read
\begin{align}
\mathcal{L}_{\rm mass} &=
-\begin{pmatrix}\bar t_L^0&\bar T_L^0\end{pmatrix}
\!\begin{pmatrix}
y_{33}^u v/\sqrt{2} & 0\\
y_{43}^u v/\sqrt{2} & M^0
\end{pmatrix}\!
\begin{pmatrix}t_R^0\\T_R^0\end{pmatrix} \nonumber\\
&-\begin{pmatrix}\bar b_L^0&\bar B_L^0\end{pmatrix}
\!\begin{pmatrix}
y_{33}^d v/\sqrt{2} & 0\\
y_{43}^d v/\sqrt{2} & M^0
\end{pmatrix}\!
\begin{pmatrix}b_R^0\\B_R^0\end{pmatrix}+{\rm h.c.},
\label{eq:L_mass}
\end{align}
where $v=246$~GeV and $M^0$ is the gauge-invariant VLQ mass, possibly generated by a singlet scalar with a large VEV.  

In the $TB$ doublet case, gauge symmetry forbids $y_{34}^{u,d}$, leaving only $y_{33}^{u,d}$ and $y_{43}^{u,d}$ nonzero. The physical states are obtained through bi-unitary transformations
$U_L^q\,\mathcal{M}^q\,(U_R^q)^\dagger = \mathcal{M}^q_{\rm diag}$,
leading to mixing angles
\begin{equation}
\tan\theta_L^u \!\simeq\! \frac{m_t}{m_T}\tan\theta_R^u,\qquad
\tan\theta_L^d \!\simeq\! \frac{m_b}{m_B}\tan\theta_R^d.
\end{equation}

\vspace{3pt}
\noindent\textbf{Decay into charged Higgs bosons.}
Within this setup, the vector-like top quark ($T$) predominantly decays as $T\!\to\! H^+ b$ whenever the channel is kinematically open, providing a direct probe of the charged Higgs state~\cite{Benbrik:2022kpo}.  
The partial width can be expressed as
\begin{align}
\Gamma(T\!\to\! H^+b)
&=\frac{g^2}{64\pi}\frac{m_T}{M_W^2}
\lambda^{1/2}(m_T,m_b,M_{H^\pm}) \nonumber\\
&\times\!\Big[(|Z_{Tb}^L|^2\!\cot^2\!\beta+|Z_{Tb}^R|^2\!\tan^2\!\beta)
(1+r_b^2-r_{H^\pm}^2)\nonumber\\
&\hspace{10pt}+\,4r_b\,\Re(Z_{Tb}^L Z_{Tb}^{R*})\Big],
\label{eq:GammaT}
\end{align}
with $r_x=m_x/m_T$ and $\lambda(x,y,z)=(x^4+y^4+z^4-2x^2y^2-2x^2z^2-2y^2z^2)$.  
The couplings $Z_{Tb}^{L,R}$ depend on the mixing angles in the up- and down-type sectors:
\begin{align}
Z_{Tb}^L &= c_L^d s_L^u e^{-i\phi_u}
+(s_L^{u\,2}-s_R^{u\,2})\frac{s_L^d}{c_L^u}, \nonumber\\
Z_{Tb}^R &= \frac{m_b}{m_T}\!\left[
c_L^d s_L^u +(s_R^{d\,2}-s_L^{d\,2})\frac{c_L^u}{s_L^d}
\right].
\end{align}

\subsection{Theoretical and Experimental Bounds}

The viability of each parameter point is tested against the following theoretical and experimental constraints:

\begin{itemize}
	
	\item \textbf{Unitarity:}  The eigenvalues of the $S$-wave scattering matrix for scalar-scalar, scalar-gauge, and gauge-gauge channels are required to satisfy perturbative unitarity at high energies~\cite{Kanemura:1993hm}.
	
	\item \textbf{Perturbativity}:  The quartic couplings of the scalar potential obey $|\lambda_i|<8\pi$ ($i=1,\dots,5$)~\cite{Branco:2011iw}.
	
	\item \textbf{Vacuum stability}  The scalar potential is required to be bounded from below in all field directions, which implies~\cite{Barroso:2013awa,Deshpande:1977rw}
	\begin{align}
	\lambda_1>0,\quad
	\lambda_2>0,\quad
	\lambda_3>-\sqrt{\lambda_1\lambda_2},\nonumber\\
	\lambda_3+\lambda_4-|\lambda_5|>-\sqrt{\lambda_1\lambda_2}.
	\end{align}
	
	\item \textbf{Constraints from EWPOs:}  The oblique parameters $S$ and $T$, including contributions from both the 2HDM-II and VLQ sectors, are required to be consistent with the global fit at the 95\%~CL~\cite{Molewski:2021ogs}:
	\begin{align}
	S = 0.05 \pm 0.08,\quad
	T = 0.09 \pm 0.07,\nonumber\\
	\rho_{S,T}=0.92
	\qquad (\text{for }U=0).
	\end{align}
	Unitarity, perturbativity, vacuum stability, and EWPO constraints are evaluated using \texttt{2HDMC-1.8.0}~\cite{Eriksson:2009ws}, modified to incorporate the VLQ interactions and the analytic $S_{\mathrm{VLQ}}$ and $T_{\mathrm{VLQ}}$ expressions of Ref.~\cite{Arhrib:2024tzm}.\footnote{The code has been extended to include new VLQ couplings and the analytic expressions for the VLQ contributions to the oblique parameters, following Ref.~\cite{Arhrib:2024tzm}.}
	
	\item \textbf{SM-like Higgs boson constraints:}  Compatibility with Higgs signal-strength measurements is imposed using \texttt{HiggsSignal-3}~\cite{Bechtle:2020pkv,Bechtle:2020uwn} via \texttt{HiggsTools-1.2}~\cite{Bahl:2022igd}, requiring
	\[
	\Delta\chi^2 = \chi^2-\chi^2_{\rm min} \le 6.18
	\]
	at the 95\%~CL.
	
	\item \textbf{Direct searches for additional scalars:}  Exclusion limits from LEP, Tevatron, and the LHC are applied at the 95\%~CL using \texttt{HiggsBounds-6}~\cite{Bechtle:2008jh,Bechtle:2011sb,Bechtle:2013wla,Bechtle:2015pma} through \texttt{HiggsTools}, including the latest searches for neutral and charged Higgs bosons.
	
	In addition, the loop effects of VLQs in $h\!\to\! gg$ and $h\!\to\!\gamma\gamma$ were analysed in Ref.~\cite{Arhrib:2024tzm}. Owing to the decoupling behaviour at large VLQ masses and the mixing constraints ($s^{u,d}_{L,R}\!\sim\!0.2$), these corrections remain moderate: $\mathcal{BR}(h\!\to\! gg)$ and $\mathcal{BR}(h\!\to\!\gamma\gamma)$ change by up to $\sim10\%$ and $\sim3\%$, respectively, predominantly through modifications of the $ht\bar t$ coupling, while direct $hT\bar T$ contributions are negligible.
	
	\item \textbf{Constraints from $b\!\to\! s\gamma$:}  In the Type-II 2HDM, this observable typically implies $m_{H^\pm}\gtrsim580$~GeV~\cite{Benbrik:2022kpo}. The presence of VLQs can partially relax this bound via additional loop contributions, however, EWPO constraints restrict the mixing angles, limiting the extent of this relaxation. In practice, viable regions arise with~\cite{Benbrik:2022kpo}
	\[
	m_{H^\pm}\!\simeq\!580~\text{GeV} \quad\text{(2HDM+$T$ singlet)}, 
	\]
	and
	\[
	m_{H^\pm}\!\gtrsim\!360~\text{GeV} \quad\text{(2HDM-II+$TB$ doublet)},
	\]
	with lighter masses possible only in restricted corners of parameter space.
	
	\item \textbf{LHC direct searches for VLQs:}  Present limits mainly target the SM decay channels $T\!\to\!Wb$, $Zt$, and $ht$. In our framework, additional modes such as $T\!\to\!H^\pm b$, $Ht$, and $At$ may alter these constraints. We incorporate the latest ATLAS and CMS results~\cite{Benbrik:2024fku} and retain only points satisfying
	\[
	r=\frac{\sigma_{\rm theo}}{\sigma_{\rm obs}^{\rm LHC}} < 1,
	\]
	corresponding to compatibility with the 95\%~CL exclusion bounds.
\end{itemize}

\section{Signal Strategy and Parameter Scan}
\label{sec-A}

Our analysis is carried out in the 2HDM+$TB$ framework, which features an enhanced charged-Higgs yield through the dominant decay mode $T\!\to\!H^+b$. In the high-mass regime ($m_T>1$~TeV), this channel reaches a branching fraction close to unity, in contrast to other VLQ representations where several decay modes compete~\cite{Arhrib:2024tzm,Benbrik:2022kpo}.

\begin{figure}[htbp!]
	\centering
	\includegraphics[height=7.5cm,width=8cm]{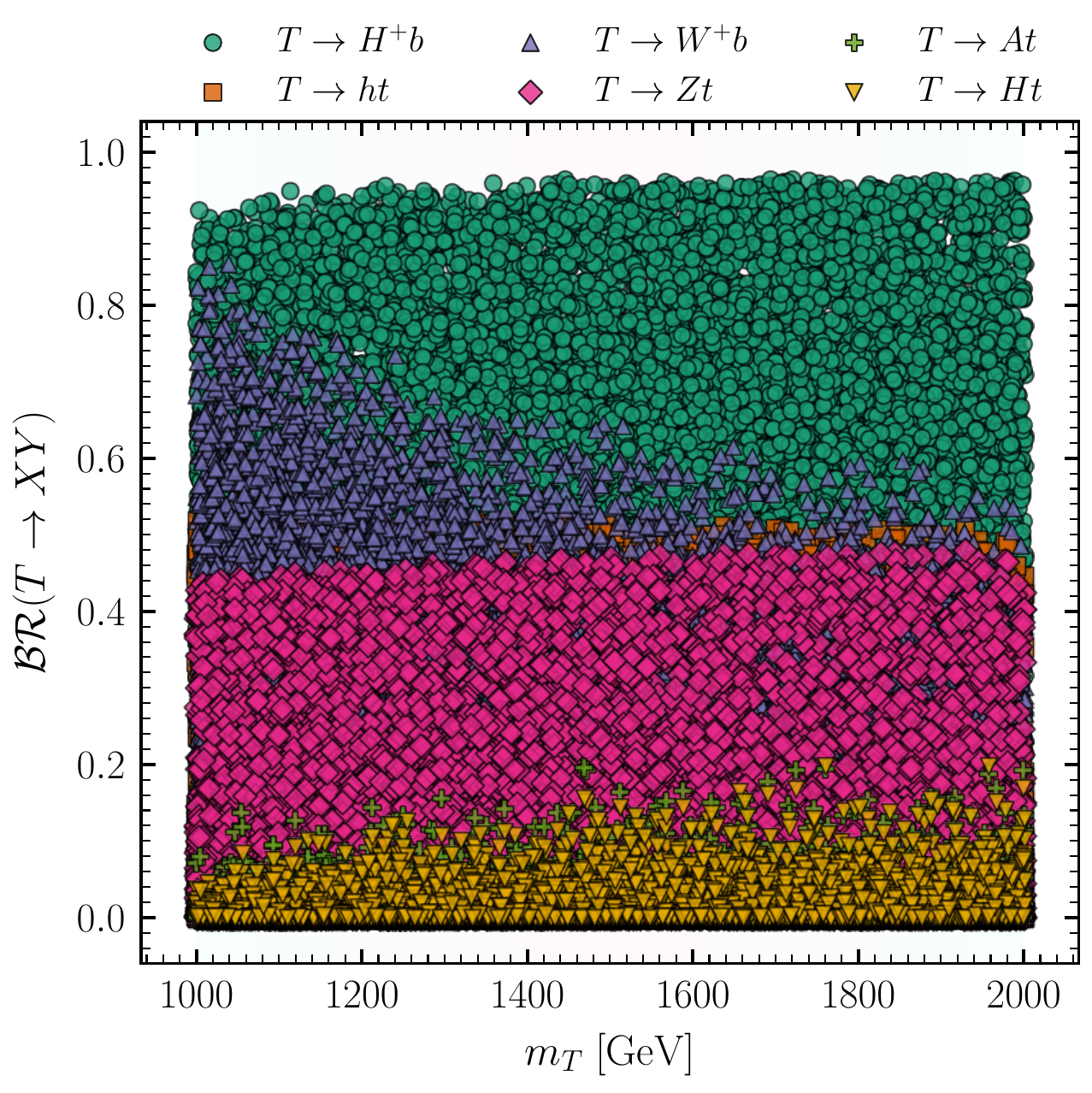}
	\caption{$\mathcal{BR}(T \to XY)$ as a function of $m_T$, with
		$XY = H^+ b$ \textnormal{(teal green)},
		$ht$ \textnormal{(orange)},
		$W^+ b$ \textnormal{(violet)},
		$Z t$ \textnormal{(magenta)},
		$A t$ \textnormal{(green)},
		and $H t$ \textnormal{(golden yellow)}.}
	\label{Fig1}
\end{figure}

A systematic scan over the parameter space is performed within the ranges listed in Table~\ref{tab:Tab1}, retaining only points consistent with all theoretical and experimental constraints discussed previously. As illustrated in Fig.~\ref{Fig1}, the decay $T\!\to\!H^+b$ dominates the VLT decay pattern throughout most of the scanned mass region for $m_T>1$~TeV.

\begin{table}[ht!]
	\centering
	\renewcommand{\arraystretch}{1}
	\setlength{\tabcolsep}{0.15\columnwidth}
	\begin{tabular}{cc}\noalign{\hrule height 0.9pt}\noalign{\vspace{1.35pt}}\noalign{\hrule height 0.4pt}
		Parameter  & Range \\
		\noalign{\hrule height 0.9pt}
		$m_h$   & $125.09$ GeV \\
		$m_A$  & [$400$, $800$] GeV \\
		$m_H$  & [$400$, $800$] GeV \\
		$m_{H^\pm}$  & [$400$, $800$] GeV \\
		$t_\beta$ & [$1$, $20$] \\
		$s_{\beta-\alpha}$ & $1$ \\
		$m^2_{12}$ & $m_A^2 s_\beta c_\beta$ \\
		$m_T$   & [$1000$, $2000$] GeV \\
		$s_L^{u}$  & [$-0.8$, $0.8$] \\
		$s_R^{d}$  & [$-0.8$, $0.8$] \\
		\noalign{\hrule height 0.4pt}\noalign{\vspace{1.35pt}}\noalign{\hrule height 0.9pt}
	\end{tabular}
	\caption{Parameter ranges explored for the 2HDM+$TB$ framework.}
	\label{tab:Tab1}
\end{table}

To optimise the signal yield and maintain sizable values of both
$\mathcal{BR}(T\!\to\!H^+b)$ and $\mathcal{BR}(H^+\!\to\!tb)$ across the accessible VLT mass range, we select four representative benchmark points with
$m_T = 1209.30$, $1452.71$, $1646.60$, and $1877.38$~GeV. The corresponding input parameters and branching fractions are reported in Table~\ref{tab:Tab2}.

\begin{table}[htp!]
	\centering
	\setlength{\tabcolsep}{5.8pt}
	\renewcommand{\arraystretch}{01}
	\begin{adjustbox}{max width=\textwidth}
		\begin{tabular}{lcccc}
			\noalign{\hrule height 0.9pt}\noalign{\vspace{1.35pt}}\noalign{\hrule height 0.4pt}
			Parameters & BP$_1$ & BP$_2$ & BP$_3$ & BP$_4$ \\
		\noalign{\hrule height 0.9pt}
			$m_h$   & 125.09 & 125.09 & 125.09 & 125.09 \\
			$m_H$ & 640.62 & 640.99 & 655.93 & 651.30 \\
			$m_A$  & 639.27 & 640.51 & 655.27 & 648.64 \\
			$m_{H\pm}$ & 642.49 & 648.62 & 647.40 & 671.88 \\
			$t_\beta$ & 5.51 & 5.04 & 4.22 & 5.22 \\
			$s_{\beta-\alpha}$ & 1 & 1 & 1 & 1 \\
			$m_T$ & 1209.30 & 1452.71 & 1646.60 & 1877.38 \\
			$m_B$ & 1217.69 & 1468.14 & 1662.59 & 1888.71 \\
			$s^u_L$ & 0.00008 & 0.00659 & $-0.00706$ & $-0.00221$ \\
			$s^d_L$ & 0.00046 & $-0.00051$ & $-0.00045$ & $-0.00028$ \\
			$s^u_R$ & 0.00056 & 0.05536 & $-0.06717$ & $-0.02405$ \\
			$s^d_R$ & 0.11718 & $-0.15450$ & $-0.15338$ & $-0.11194$ \\
			\noalign{\hrule height 0.9pt}
			\multicolumn{5}{c}{$\mathcal{BR}$ in \%} \\
			\noalign{\hrule height 0.9pt}
			${\cal BR}(T\to H^+b)$  & 93.98 & 93.49 & 91.43 & 95.19 \\
			${\cal BR}(H^+\to tb)$  & 95.63 & 96.86 & 98.33 & 96.42 \\
			\noalign{\hrule height 0.4pt}\noalign{\vspace{1.35pt}}\noalign{\hrule height 0.9pt}
		\end{tabular}
	\end{adjustbox}
	\caption{Benchmark points (BPs) for the 2HDM+$TB$ framework. Masses are in GeV.}
	\label{tab:Tab2}
\end{table}

\label{sec3}

\begin{figure}[htbp!]
	
		\includegraphics[height=5.25cm,width=8cm]{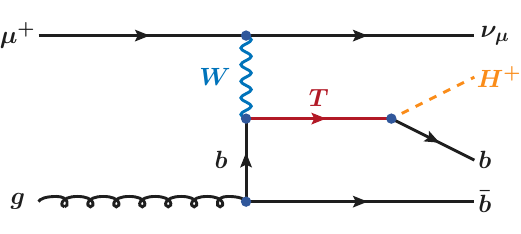}
	\caption{Representative Feynman diagram for the single production of a VLT quark, followed by its decay into a charged Higgs boson and a bottom quark.}
	\label{diag}
\end{figure}

We concentrate on the single production of the vector-like top partner at a future $\mu p$ collider with
\begin{equation}
\sqrt{s}=9.16~\text{TeV},\qquad
E_p=7~\text{TeV},\qquad
E_\mu=3~\text{TeV},
\end{equation}
and analyse the charged-Higgs-mediated channel
\begin{equation}
\mu^+ p \to \nu_{\mu}\,\bar b T
\to \nu_{\mu}\,\bar b H^+ b
\to \nu_{\mu}\,\bar b t\bar b b
\to \nu_{\mu}\,\bar b\bar b b b\,\ell^+\nu_\ell,
\end{equation}
which leads to a final state with four $b$-jets, a single charged lepton, and missing transverse energy.
For comparison, a related study of the same $T\!\to\!H^+b$ decay chain in single-$T$ production at the 14~TeV HL-LHC is presented in Ref.~\cite{Benbrik:2023xlo}.

\subsection{Event generation, detector simulation, and selection}
\label{sec:detector}

The 2HDM-II+VLQ setup is implemented in \texttt{FeynRules}~\cite{Alloul:2013bka} and exported in UFO format~\cite{Degrande:2011ua}. Signal and background events are generated at parton level with \texttt{MadGraph5\_aMC@NLO~v2.9.14}~\cite{Alwall:2014hca} and subsequently processed with \texttt{Pythia~8.30}~\cite{Sjostrand:2014zea} for parton showering and hadronization. Detector effects are simulated using \texttt{Delphes~3.5.0}~\cite{deFavereau:2013fsa} with an LHeC-like detector configuration, and jets are reconstructed with the anti-$k_t$ algorithm~\cite{Cacciari:2008gp} with radius parameter $R=0.4$. Parton densities are described by the \texttt{NN23LO1} PDF set~\cite{NNPDF:2014otw}. The final event selection and analysis are performed within the \texttt{MadAnalysis5} framework~\cite{Conte:2013mea}.

At generator level, minimal acceptance requirements are applied to both signal and background samples:
\begin{itemize}
	\item $p_T^{b/j} > 15~\text{GeV}$ for $b$-jets and light jets,
	\item $|\eta_\ell| < 2.5$ for charged leptons,
	\item $\Delta R(x,y) > 0.4$ for all pairs $x,y = j,b,\ell$.
\end{itemize}

\begin{figure*}[htpb!]
	\begin{center}
		\includegraphics[height=18cm,width=15cm]{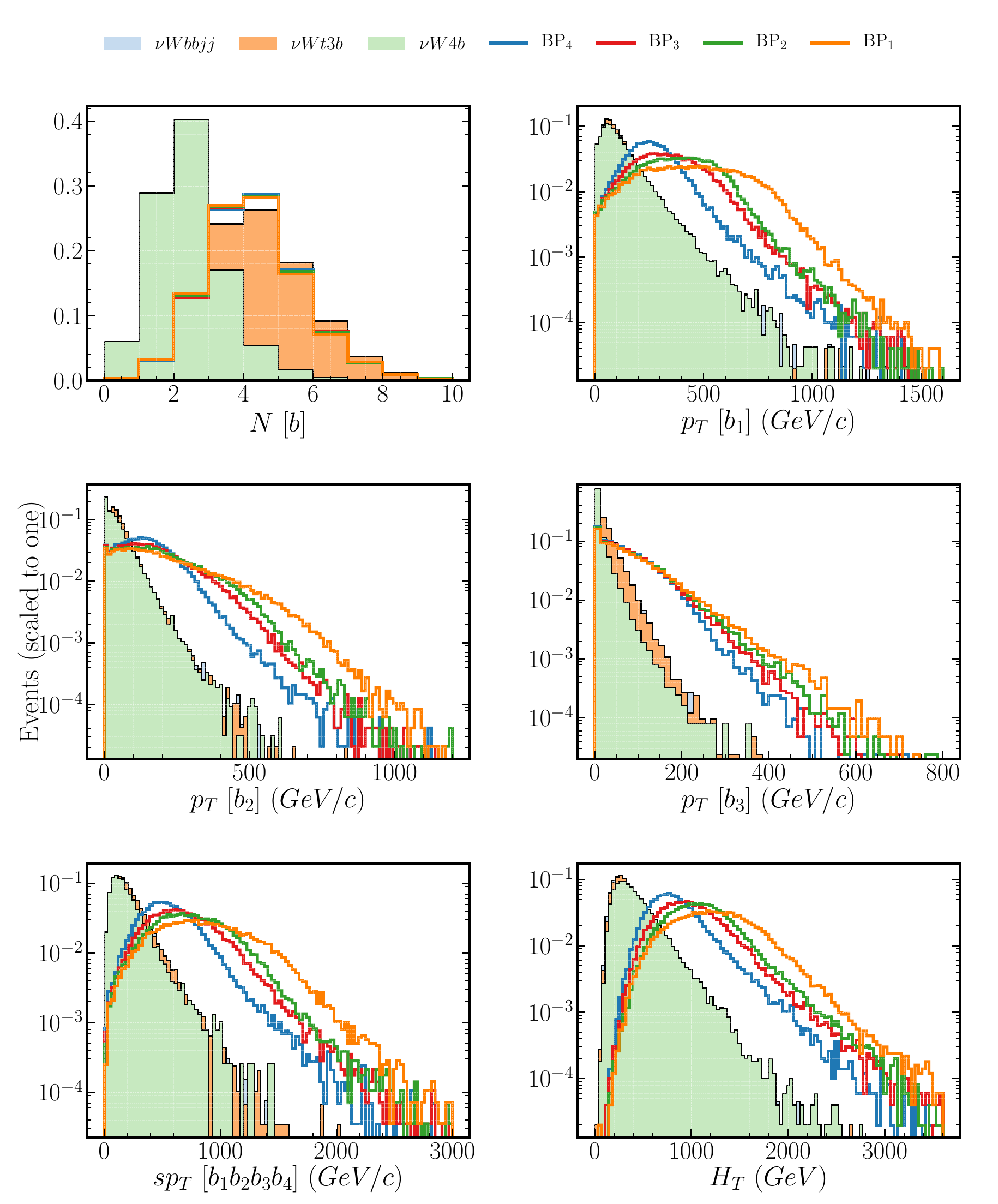}
		\caption{Normalized distributions of $N[b]$, $p_T[b_1]$, $p_T[b_2]$, $p_T[b_3]$, $sp_T[b_1b_2b_3b_4]$, and $H_T$ for the four benchmark points and the SM backgrounds at $\sqrt{s}=9.16$~TeV.}
		\label{fig:kinematics}
	\end{center}
\end{figure*}

The key kinematic observables used to define the cut strategy are displayed in Fig.~\ref{fig:kinematics}, where we show, for the four benchmark points (BPs) and the dominant SM backgrounds, the normalized distributions of the $b$-jet multiplicity, the transverse momenta of the leading $b$-jets, the scalar sum of their transverse momenta, and the global event variable $H_T$.

Guided by these observables, we design the following sequence of analysis cuts, summarized in Tab.~\ref{cutflow4}:

\begin{table}[htpb!]
	\centering
	\setlength{\tabcolsep}{10pt}
	\renewcommand{\arraystretch}{0.9}
	\begin{adjustbox}{max width=\textwidth}
		\begin{tabular}{cc} 
			\noalign{\hrule height 0.9pt}\noalign{\vspace{1.35pt}}\noalign{\hrule height 0.4pt}
			Cut & Definition \\  
			\noalign{\hrule height 0.9pt}
			Cut 1 & $N(b) \geq 4$, \quad $N(\ell^+) = 1$ \\
			Cut 2 & $p_T(\ell^+) > 50~\text{GeV}$, \quad $\slashed{E}_T > 50~\text{GeV}$ \\
			Cut 3 & $|\eta_{\ell^+}| < 1.2$, \quad $|\eta_{b_1}| < 1.2$, \quad $|\eta_{b_2}| < 1.2$ \\
			Cut 4 & $p_T(b_1) > 240~\text{GeV}$, \quad $p_T(b_2) > 100~\text{GeV}$, \\
			& $p_T(b_3) > 60~\text{GeV}$, \quad $p_T(b_4) > 50~\text{GeV}$ \\
			Cut 5 & $sP_T > 600~\text{GeV}$ \\
			Cut 6 & $H_T > 980~\text{GeV}$ \\
			\noalign{\hrule height 0.4pt}\noalign{\vspace{1.35pt}}\noalign{\hrule height 0.9pt}
		\end{tabular}
	\end{adjustbox}
	\caption{Sequence of cuts used in the analysis of signal and background events at $\sqrt{s}=9.16$~TeV.}
	\label{cutflow4}
\end{table}

The corresponding cut flow for the signal and the dominant SM backgrounds is reported in Tab.~\ref{cutflow2}.  
Cross sections are given in fb after each selection step, together with the overall efficiencies.

\begin{table*}[htpb!]
	\centering
	\setlength{\tabcolsep}{8pt}
	\renewcommand{\arraystretch}{1.05}
	\begin{adjustbox}{max width=1\textwidth}	
		\begin{tabular}{p{1.7cm}<{\centering} p{1cm}<{\centering} p{1cm}<{\centering} p{1cm}<{\centering} p{0.2cm}<{\centering}  p{1.6cm}<{\centering} p{1.6cm}<{\centering} p{1.4cm}<{\centering}}
			\noalign{\hrule height 0.9pt}\noalign{\vspace{1.35pt}}\noalign{\hrule height 0.4pt}
			\multirow{2}{*}{Cuts} & \multicolumn{4}{c}{Signal} & \multicolumn{3}{c}{Backgrounds} \\ 
			\cline{2-5} \cline{6-8}
			& BP1 & BP2 & BP3 & BP4 & $\nu_{\mu}W^+4b$ & $\nu_{\mu}W^+3bt$ & $\nu_{\mu}W^+2b2j$ \\
			\noalign{\hrule height 0.9pt}
			Basic & 11.50 & 4.82 & 2.34 & 1.44 & 18.50 & 17.91 & 2260.40 \\
			Cut 1 & 5.06  & 2.07 & 1.00 & 0.60 & 8.64  & 8.36  & 134.80  \\
			Cut 2 & 2.71  & 1.31 & 0.68 & 0.40 & 2.79  & 2.89  & 51.18   \\
			Cut 3 & 1.22  & 0.63 & 0.35 & 0.22 & 0.887 & 0.898 & 14.38   \\
			Cut 4 & 0.266 & 0.170 & 0.099 & 0.067 & 0.013 & 0.014 & 0.136  \\
			Cut 5 & 0.247 & 0.164 & 0.097 & 0.066 & 0.010 & 0.011 & 0.136  \\
			Cut 6 & 0.153 & 0.128 & 0.082 & 0.060 & 0.0025 & 0.0050 & 0 \\
			Eff. [\%] & 1.33 & 2.64 & 3.51 & 4.19 & $1.3\times10^{-4}$ & $2.8\times10^{-4}$ & 0 \\
			\noalign{\hrule height 0.4pt}\noalign{\vspace{1.35pt}}\noalign{\hrule height 0.9pt}
		\end{tabular}
	\end{adjustbox}
	\caption{Cut flow of the cross sections (in fb) for the signal and SM backgrounds at a $\mu p$ collider with $\sqrt{s}=9.16$~TeV for the four benchmark points.}
	\label{cutflow2}
\end{table*}

\subsection{Statistical interpretation and discovery reach}

To quantify the sensitivity of the proposed search, we adopt the median-significance formalism of Ref.~\cite{Cowan:2010js}.  
Including a fractional systematic uncertainty $\delta$ on the background normalisation, the discovery (or exclusion) significance is given by
\begin{widetext}
	\begin{align}
	\mathcal{Z} &= \sqrt{2\left[(s+b)\ln\left(\frac{(s+b)(1+\delta^2 b)}{b+\delta^2 b (s+b)}\right)
		-\frac{1}{\delta^2}\ln\left(1+\delta^2\frac{s}{1+\delta^2 b}\right)\right]},
	\end{align}
\end{widetext}
where $s$ and $b$ denote the expected number of signal and background events after all selection cuts.  
In the limit $\delta \to 0$, this expression reduces to the usual form
\begin{align}
\mathcal{Z} &= \sqrt{2\left[(s+b)\ln\left(1+\frac{s}{b}\right)-s\right]}.
\end{align}

Using the cross sections in Tab.~\ref{cutflow2}, we compute the discovery significance for the four BPs at $\sqrt{s}=9.16$~TeV, for integrated luminosities of 100 and 234~fb$^{-1}$ and for several representative values of the systematic uncertainty.  
The results are summarized in Tab.~\ref{tab:significance_sys}.

\begin{table}[htpb!]
	\centering
	\setlength{\tabcolsep}{7pt}
	\renewcommand{\arraystretch}{1.1}
	\begin{tabular}{|c|c|cccc|}\noalign{\hrule height 0.9pt}\noalign{\vspace{1.35pt}}\noalign{\hrule height 0.4pt}

		$\mathcal{L}$ [fb$^{-1}$] & Systematic & BP$_1$ & BP$_2$ & BP$_3$ & BP$_4$ \\
	\noalign{\hrule height 0.9pt}
		\multirow{4}{*}{100} 
		& 0\%  & 8.23 & 7.27 & 5.29 & 4.20 \\
		& 10\% & 7.48 & 6.60 & 4.79 & 3.84 \\
		& 20\% & 6.23 & 5.48 & 4.04 & 3.26 \\
		& 30\% & 5.28 & 4.64 & 3.43 & 2.78 \\
	\noalign{\hrule height 0.9pt}
		\multirow{4}{*}{234} 
		& 0\%  & 12.59 & 11.12 & 8.09 & 6.43 \\
		& 10\% & 11.39 & 10.09 & 7.33 & 5.86 \\
		& 20\% & 9.47 & 8.39 & 6.08 & 4.87 \\
		& 30\% & 8.02 & 7.11 & 5.15 & 4.13 \\
\noalign{\hrule height 0.4pt}\noalign{\vspace{1.35pt}}\noalign{\hrule height 0.9pt}

	\end{tabular}
	\caption{Discovery significances for the four benchmark points at $\sqrt{s}=9.16$~TeV for $\mathcal{L}=100$ and 234~fb$^{-1}$, including systematic uncertainties of 0\%, 10\%, 20\%, and 30\%.}
	\label{tab:significance_sys}
\end{table}

\begin{figure*}[htbp!]
	\centering
	\includegraphics[height=7.25cm,width=13cm]{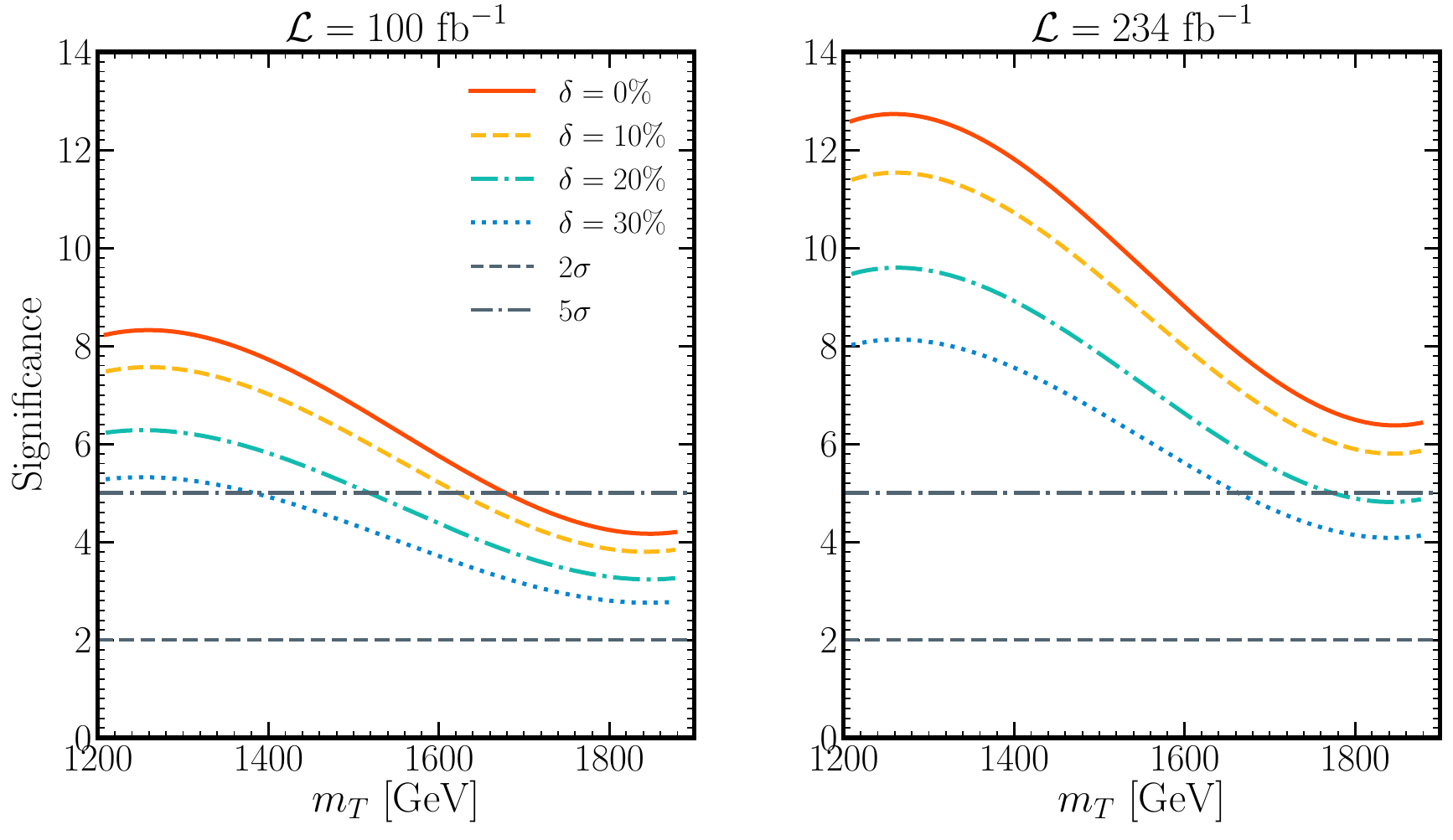}
	\caption{Discovery significance as a function of $m_T$ for different values of the systematic uncertainty $\delta$ and integrated luminosities $\mathcal{L}=100$ and $234$~fb$^{-1}$ at $\sqrt{s}=9.16$~TeV. The parameters are fixed to $m_{H^\pm}\simeq 600$~GeV with $m_H\simeq m_A\simeq m_{H^\pm}$, $\tan\beta = 5$, $s_R^u = 0.05$, and $s_R^d = 0.11$. All shown points satisfy the theoretical and experimental constraints discussed in the text.}
	\label{fig_sig}
\end{figure*}

The results collected in Table~\ref{tab:significance_sys} and summarised in Fig.~\ref{fig_sig} exhibit a clear dependence on the VLT mass. The largest sensitivity is obtained for BP$_1$ ($m_T\simeq 1.21$~TeV), while BP$_2$ ($m_T\simeq 1.45$~TeV) and BP$_3$ ($m_T\simeq 1.65$~TeV) display a progressively reduced significance, consistent with the decrease of the production cross section at larger $m_T$. These three benchmarks nevertheless remain at, or very close to, discovery level for $\mathcal{L}=100$~fb$^{-1}$ under moderate systematic uncertainties. For the heaviest configuration, BP$_4$ ($m_T\simeq 1.88$~TeV), the behaviour of the interpolated curves in Fig.~\ref{fig_sig} shows that discovery requires higher luminosity: at $\mathcal{L}=234$~fb$^{-1}$ all benchmarks exceed the $5\sigma$ threshold for $\delta\lesssim20\%$, with a comfortable margin for the lighter mass points.

\section{Discussion and Summary}
\label{conclusion}

We have investigated the discovery prospects for a vector-like top partner ($T$) in the 2HDM-II+$TB$ framework at a future high-energy $\mu p$ collider, focusing on the charged-Higgs mediated decay chain $T\!\to\!H^+b \to t\bar{b}\,b$. In this scenario, the single-production topology provides a clean experimental signature characterised by four $b$-jets, one isolated charged lepton, and missing transverse energy. A comprehensive parameter scan, subject to theoretical and experimental constraints, shows that in the high-mass regime the decay $T\!\to\!H^+b$ dominates the VLT branching pattern, while $H^+\!\to\!tb$ remains the leading charged-Higgs decay mode.

A detailed detector-level simulation was performed for four representative benchmark points with $m_T\simeq 1.2$-$1.9$~TeV. The optimized selection strategy yields strong background suppression and high signal purity, enabling significant sensitivity across the entire benchmark set. The resulting discovery significances exhibit a clear mass-ordered behaviour: lighter benchmarks achieve discovery level already at $\mathcal{L}=100$~fb$^{-1}$, while the heaviest scenario requires larger statistics. At $\mathcal{L}=234$~fb$^{-1}$, all benchmarks surpass the $5\sigma$ threshold, and discovery remains robust even under sizable systematic uncertainties.

Overall, our findings demonstrate that a future $\mu p$ collider offers excellent sensitivity to charged Higgs bosons produced via $T\!\to\!H^+b$, extending the discovery reach for vector-like top masses up to nearly $2$~TeV and providing a powerful probe of extended Higgs sectors with vector-like quarks.

\section*{ACKNOWLEDGMENTS}
M. Boukidi acknowledges the support of the Narodowe Centrum Nauki under OPUS Grant No. 2023/49/B/ST2/03862 as well as the use of the PALMA II high-performance computing cluster at the University of Münster, subsidised by the DFG (INST 211/667-1), in completing this work. The authors are grateful for the technical support provided by CNRST/HPC-MARWAN.
\bibliographystyle{JHEP}
\bibliography{main} 
\end{document}